\documentclass[11pt,twoside]{article}
\usepackage{asp2004}
\usepackage{epsf}
\usepackage{graphics}
\usepackage{lscape}
\def\edcomment#1{\iffalse\marginpar{\raggedright\sl#1\/}\else\relax\fi}
\reversemarginpar

\begin{document}
\title{Radio evidence on the mass loss bi-stability jump}
 \author{P. Benaglia}
   \affil{Instituto Arg. de Radioastronom\'{\i}a, C.C.5, (1894) V. Elisa, Argentina}
 \author{J. S. Vink}
   \affil{Keele University, Astrophysics, Lennard-Jones Lab, ST5 5BG, UK}
 \author{J. Mart\'{\i}}
   \affil{Dep. de F\'{\i}s., U. de Ja\'en, C. Las Lagunillas, E. A3, 23071 Ja\'en, Spain}
 \author{J. Ma\'{\i}z-Apell\'aniz}
    \affil{STScI, 3700 San Martin Drive, Baltimore, MD 21218, U.S.A.}
 \author{B. Koribalski \& S. Johnston}
    \affil{ATNF, CSIRO, PO Box 76, Epping, NSW 1710, Australia}

\begin{abstract}
We present the results of a first campaign of radio continuum
observations toward supergiants with spectral types in the range
O8 - B3. Three targets out of twelve were detected with the ATCA
and the VLA. The measured fluxes and the upper limits yielded
values of the stellar  mass-loss rates $\dot{M}$ and the wind
efficiencies. The comparison of predicted and derived values of
$\dot{M}$ shows a good agreement.
\end{abstract}

\vspace{0.5cm}


Lamers et al. (1995) showed that for massive, early-type stars, the ratio
between the measured terminal velocity and the escape velocity
drops from 2.6 to 1.3, at stellar effective temperatures ($T_{\rm
eff}$) around 25 kK. This so-called bi-stability jump (BSJ) was
studied later by Vink et al. (1999), who found it to be due to the change
in the ionization of the Fe lines that drive the wind. The authors
also predict a jump in the mass-loss rate {\it by a factor of
five}, for stars at the same luminosity, but this effect has to be
proved observationally.

For a thermal stellar wind, the mass loss rate can be derived by
means of radio data; the measured continuum flux density is
proportional to  $\dot{M}$ (e.g. Panagia \& Felli 1975, Benaglia et al. 2001).

In order to study the behavior of the mass loss rate at $T_{\rm
eff}$
at both sides of the BSJ, we started a
project to gather all results of previous radio observations and
to extend the sample with new detections. Here we present the new
observations.

We have performed high resolution continuum observations of the
supergiants listed in Table 1. Eight of them were observed using
the VLA, at 8.46 GHz, for $\sim$1 h each. The rest were observed
with the ATCA at 17.8 GHz, during 4 h. HD 76341, HD 148379 and HD
154090 were detected (Figs. 1-2).

Table 1 lists the target stars, some stellar parameters adopted,
and the measured flux densities ($3\sigma$ upper limit for
non-detections).
We have derived the mass loss rate and the corresponding wind
efficiency
 $\eta = {\rm c}\dot{M}v_{\infty}/{L_*}$. The last column gives the expected mass-loss rates.
At first sight, these radio data appear in reasonable agreement
with predictions, although we note a large discrepency for HD
148379. The issue therefore warrants further investigation with a
larger sample, which is forthcoming (Benaglia et al., in prep.).


\begin{table}[!ht]
\label{table1}
\caption{Observed stars and derived mass-loss rates}
\begin{center}
{\small
\begin{tabular}{lllrrrrr}
\tableline
\noalign{\smallskip}
Star &     Sp. Class. & Tel.& $d$ & $S_{\nu}$ &  $\log$  & $\eta$ & $\log$\\
     &                &        &(kpc)&    (mJy)  & $(\dot{M})$  &  &  $(\dot{M}_{\rm p})$\\
\noalign{\smallskip} \tableline \noalign{\smallskip}

HD 42087  & B2.5 Ib &VLA& 1.2& $<$ 0.14&   $<-6.06$& $<$0.60 &
$-6.9$\\

HD 43384  &  B3 Iab  &VLA& 1.4& $<$ 0.24&  $<-5.77$ & $<$0.82 &
$-6.8$\\

HD 47432  &  O9.7 Ib &VLA& 1.7&  $<$ 0.15& $<-5.51$& $<$0.64&
$-6.0$\\

HD 76341  &  O9 Ib   &VLA& 1.9& $0.38\pm0.05$ & $-5.14$& 1.31&
$-5.7$\\

HD 112244 &  O8.5 Iab(f) & ATCA& 1.5& $<$ 0.3& $<-5.52$& $<$0.52&
$-5.6$\\

HD 148379 &  B1.5 Iape & ATCA&1.0& $0.28\pm0.05$&$-6.28$& 0.03 &
$-4.4$\\

HD 154090 &  B0.7 Ia &VLA& 1.1&   $0.23\pm0.04$&  $-5.87$& $0.20$
& $-5.5$\\

HD 156154 &  O8 Iab(f) &VLA& 2.2 & $<$ 0.15 & $<-5.35$& $<$0.69 &
$-5.5$\\

HD 157246 &  B1 Ib   &ATCA& 1.1&  $<$    0.18& $<-6.21$& $<$0.08 &
$-5.5$\\

HD 165024 &  B2 Ib   &ATCA& 0.8&  $<$ 0.18& $<-6.21$& $<$0.14 &
$-5.4$\\

HD 204172 &  B0 Ib   &VLA& 3.0& $<$ 0.08& $<-5.30$& $<$1.30 &
$-6.1$\\

BD-11 4586&  O8 Ib(f)&VLA& 1.8& $<$ 0.3& $<-5.25$& $<$0.88 &
$-5.5$\\

\noalign{\smallskip} \tableline\
\end{tabular}
}
\end{center}
\end{table}

\vspace{-0.5cm}

\begin{figure}[!ht]
\plotone{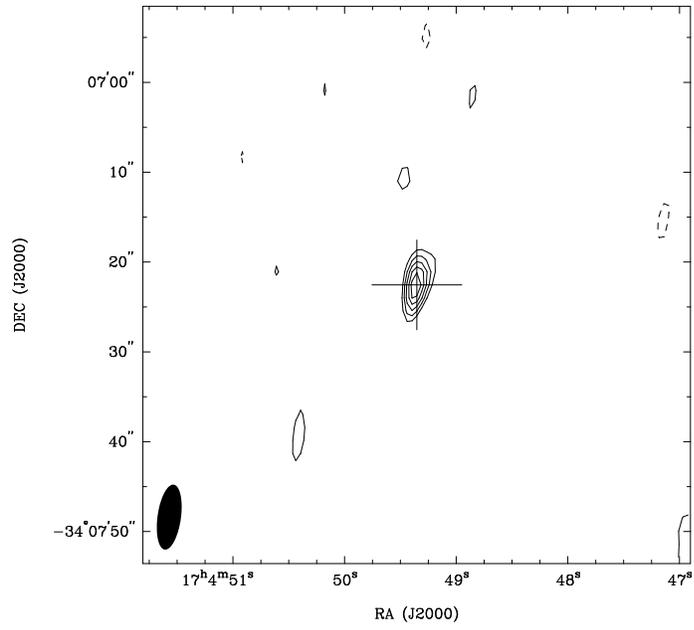}\caption{VLA image of 154090 at 8.46
GHz; levels: --0.08, 0.08 ($2\sigma$), 0.12, 0.15, 0.18. and 0.21
mJy beam$^{-1}$.}\label{figure1}
\end{figure}

\begin{figure}[!ht]
\vspace{1cm} \label{figure2}
\plottwo{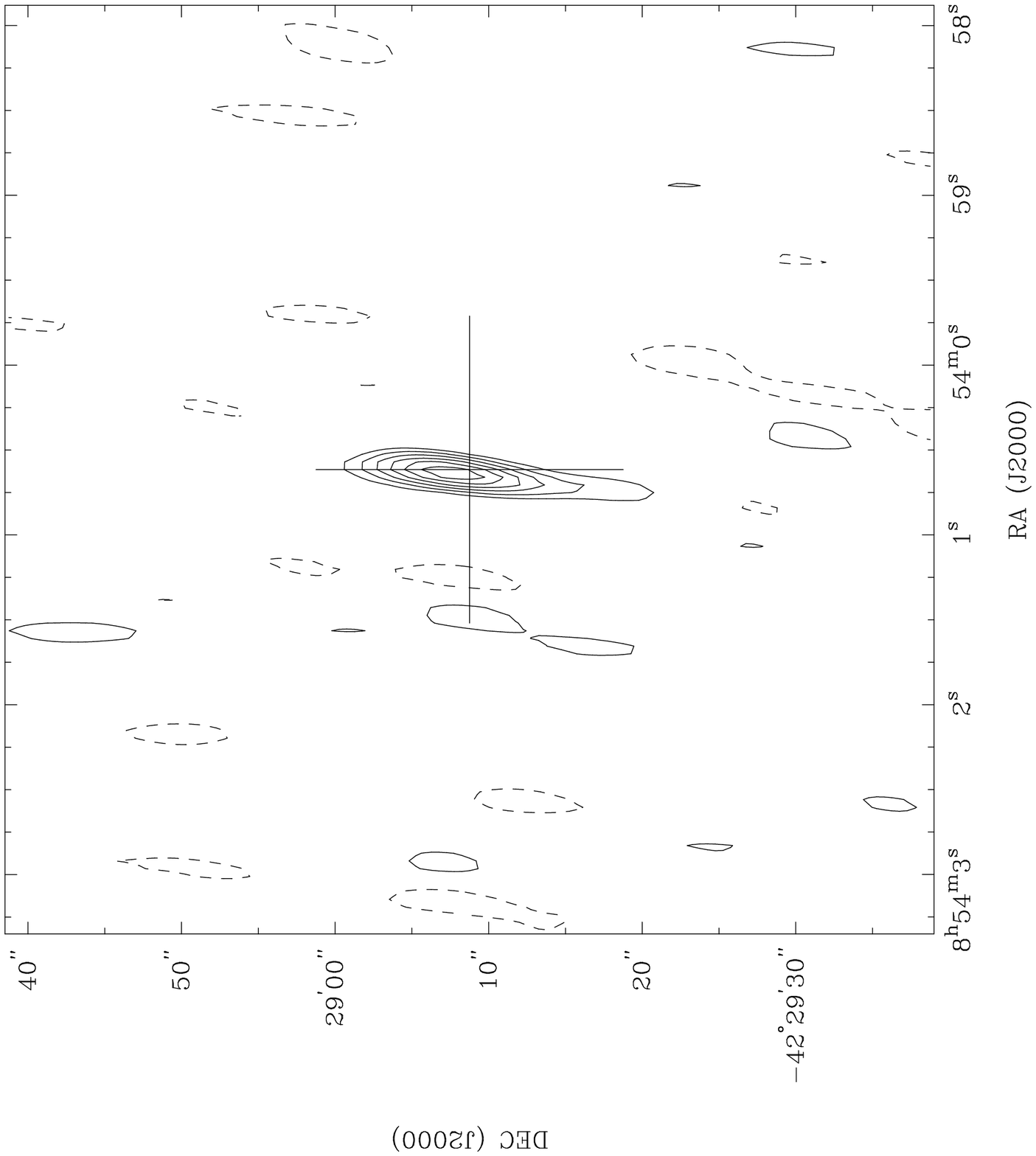}{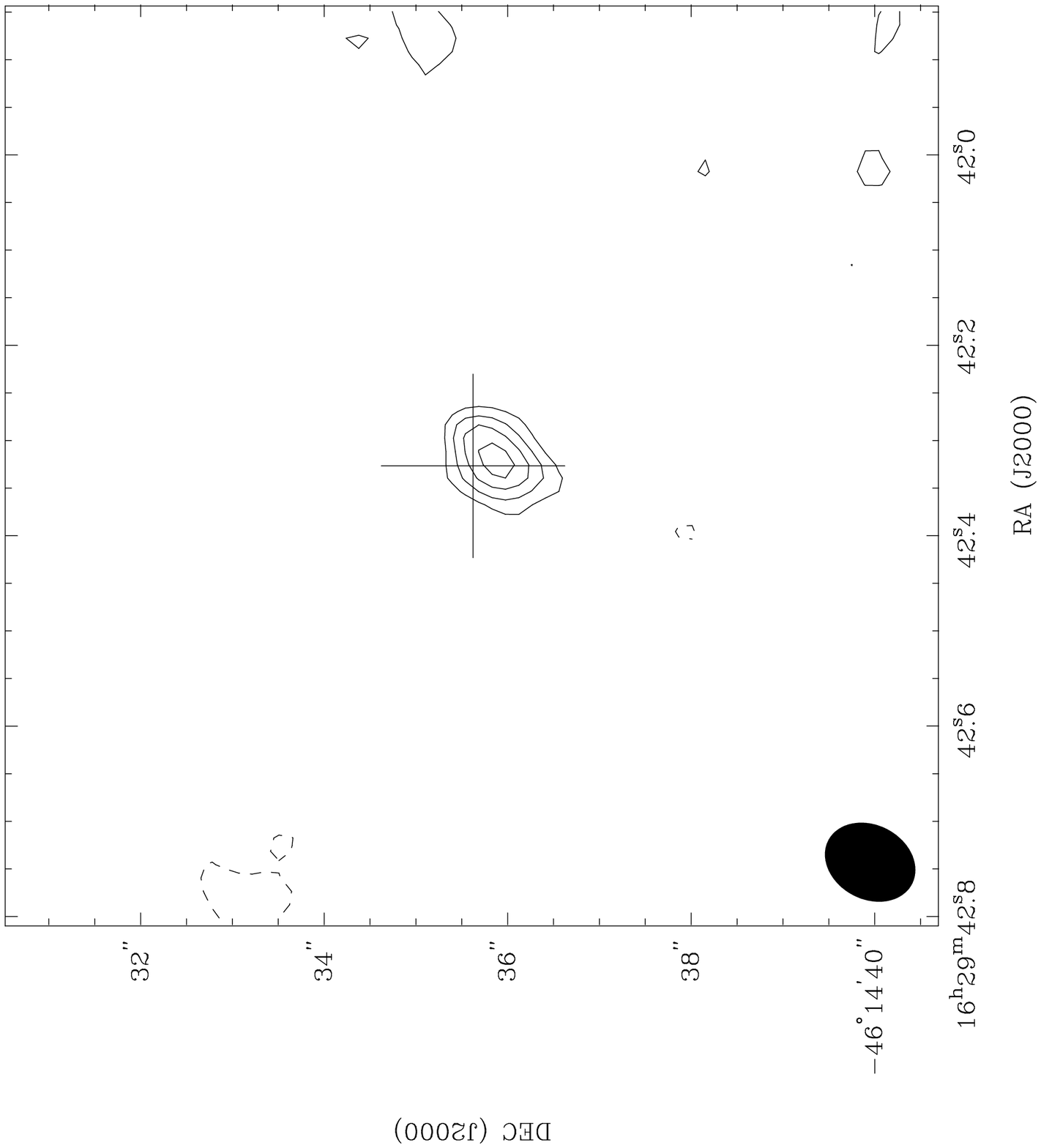}
\caption{{\sl Left}: VLA image of HD 76341 at 8.46 GHz; levels:
--0.15, 0.15 ($3\sigma$), 0.2, 0.25, 0.3 and 0.35 mJy beam$^{-1}$.
{\sl Right}: ATCA image of HD 148379 at 17.8 GHz; levels: --0.1, 0.15
($3\sigma$), 0.2, 0.25, and 0.3 mJy beam$^{-1}$.}
\end{figure}

\section*{References}

Benaglia, P., Cappa, C.E., \& Koribalski, B. 2001, A\&A, 200, 58

\noindent Lamers, H.J.G.L.M., Snow, T.P., \& Lindholm, D.M. 1995, ApJ, 455, 269

\noindent Panagia, N., \& Felli, M. 1975, A\&A, 39, 1

\noindent Vink, J.S, de Koter, A., \& Lamers, H.J.G.L.M. 1999, A\&A, 35, 181

\end{document}